\documentclass[traditabstract]{aa}
\usepackage{txfonts}
\usepackage{graphicx}
\usepackage{amssymb}
\usepackage{natbib}
\bibpunct{(}{)}{;}{a}{}{,} % to follow the A&A style

\newcommand{\eqb}{\begin{eqnarray}}
\newcommand{\eqe}{\end{eqnarray}}

\begin{document}

\title{The time-dependent one-zone hadronic model - First
principles}

\author{S. Dimitrakoudis\inst{1}
  \and A. Mastichiadis\inst{1}
    \and R.J. Protheroe\inst{2}
    \and A. Reimer\inst{3}}

\offprints{S. Dimitrakoudis, \email{sdimis@phys.uoa.gr}}

\institute{Department of Physics, University of Athens, Panepistimiopolis, GR 15783, Zografos, Greece
  \and Department of Physics (School of Chemistry and Physics), University of Adelaide, North Terrace, Adelaide, SA 5005, Australia
  \and Institut f\"ur Theoretische Physik, und Institut f\"ur Astro- und Teilchenphysik, Universit\"at Innsbruck, Austria}

\date{Received ? / Accepted ?}

\abstract {We present a time-dependent approach to the one-zone hadronic model in the case
where the photon spectrum is produced by ultrarelativistic protons 
interacting with soft photons that are produced from protons and low magnetic fields. 
Assuming that protons are injected at a certain rate in a homogeneous spherical volume containing 
a magnetic field, the evolution of the system can be described by five coupled kinetic equations, 
for protons, electrons, photons, neutrons, and neutrinos. Photopair and photopion 
interactions are modelled using the results of Monte-Carlo simulations and, in particular from 
the SOPHIA code for the latter. The coupling of energy losses and injection introduces 
a self-consistency in our approach and allows the study of 
the comparative relevancy of processes at 
various conditions, the efficiency of the conversion of proton luminosity to radiation, 
the resulting neutrino spectra, and the effects of time variability on proton injection, among other topics. 
We present some characteristic examples of the temporal behaviour of the system 
and show that this can be very different from the one exhibited by leptonic models.
Furthermore, we argue that, contrary to the wide-held belief, there are parameter
regimes where the hadronic models can become quite efficient. However,
to keep the free parameters at a minimum and facilitate an in-depth
study of the system, we have only concentrated on the case where  protons 
are injected; i.e., we did not consider the effects of a co-accelerated leptonic component.
}

\maketitle

\section{Introduction}

BL Lac objects and flat-spectrum radio quasars (commonly termed 
blazars) display a spectral energy distribution (SED) that extends from radio to $\gamma$-rays, 
in the shape of two broad humps - one in the radio to UV or X-ray frequency range and the other 
in the X-ray to $\gamma$-ray range. Both components are non-thermal and very likely originate in 
the blazar's jet, which is closely aligned with the line of sight of the observer. 
While the lower-energy hump is widely acknowledged to result from synchrotron radiation 
from relativistic electrons in the emitting region, the origin of the higher-energy 
hump is subject to differing interpretations.

According to the leptonic models, the high-energy component emerges from inverse Compton scattering of electrons 
on soft target photons. Those photons may be the product of synchrotron radiation 
of the same electrons  
\citep{maraschighisellinicelotti,BloomMarscher96,inouetakahara} 
or an indigenous population of external photons  
\citep{Dermeretal92,sikorabegelmanrees}. 
Hadronic models, on the other hand, assume that protons, which are accelerated along 
with electrons, contribute much of the high-energy component via photopion interactions 
\citep{Mannheim98,Muckeetal2003,BottcherRM2009}. 
The target photons may either be produced inside the jet via synchrotron 
emission from a co-accelerated population of electrons 
\citep{Mannheim93}, 
originate outside the jet \citep{Protheroe95,Bednarek, AtoyanDermer2001}, or be produced via 
synchrotron emission from the protons themselves {\citep{aharonian,
mueckeprotheroe}. In addition proton-proton interactions were taken into
account for the first time in the AGN case by \cite{ProtheroeKazanas1983} and were later applied to jets by \cite{Siewert2004} and \cite{Reynoso2011}.
For a recent review of the features of leptonic versus hadronic models see \cite{Boettcher2012}. Although similar models are used to study gamma ray bursts \citep{BottcherD1998, KazanasGM2002, MastiKazanas2006, AsanoInoue2007, Asanoetal2009, MastiKazanas2009} and $\gamma$-ray emitting compact binary systems \citep{Romero2003, Paredesetal2005, Romeroetal2005}, the former demand large magnetic fields, while the latter demand high proton densities, making inelastic pp-collisions more important than photohadronic interactions. However, it should be noted that both photo-pair and photo-meson production are included in recent models of compact binaries, and they are shown to be significant when proton and electron luminosities are near equipartition \citep{Vila2012}.
  The processes and range of variables used in this work are, instead, tailored to photon-dominated systems such as blazars, although they can indice similar effects in other systems, and thus are qualitatively relevant to them.

While leptonic models have often made use of time-dependent codes to solve the 
kinetic equations of electrons and photons \citep{mastikirk97, Krawczynski, BottcherChang2002,
Katarzynskietal}, 
hadronic models have been very difficult to investigate in a similar manner, owing to the greater complexity of the particle interactions involved and the resulting time-consuming nature of simulations. 
First efforts \citep{mastikirk95} used simple $\delta-$functions for the
secondary production, while 
the effects of Bethe-Heitler pair production were calculated in a time-dependent manner in 
\cite{mastiprothkirk} who used the Monte Carlo results of 
\cite{protheroejohnson} to model the process. 
In the present paper we extend this method to incorporate photopion interactions 
in detail. For this
we have required the use of the SOPHIA Monte-Carlo event generator for hadronic interactions 
\citep{SOPHIA2000}, whose results have been adapted into a time-dependent code. 

The paper is structured as follows. In \textsection 2 we present the time-dependent kinetic equations for all five particle types and show the coupling that emerges when all leptonic and hadronic processes are taken into account. In \textsection 3 we discuss the integration of the SOPHIA code's results into our model and its verifications through energy-loss tests. In \textsection 4 we present various results from monoenergetic proton injection in a magnetic field, noticing the transition in importance from photopair to photopion when the injected proton energy is increased, and the transition from linear to non-linear proton cooling when the injected proton compactness is increased; also the effects of time variability in proton injection. We move on to power-law proton injection in \textsection 5, while in \textsection 6 we show some characteristic time-dependent cases. 
Finally, in \textsection 7
we conclude with a summary and make some remarks on the results of the previous sections.

\section{The kinetic equation approach}

\subsection{Setup of the problem}

Relativistic protons in compact sources undergo
a series of processes that can be quite complicated to model.
Thus protons interact with photons creating pions, neutrons and
neutrinos (all from photopion interactions), and also electron-positron
pairs (from photopair). Since the magnetic field in the cases examined in this paper is low ($ B \leq 10~{\rm G} $) and the maximum proton Lorentz factor does not exceed $\gamma_p=10^8$, the pions and their resulting muons decay almost instantaneously, as compared with
any other relevant timescale of the problem, creating photons and
more electrons, positrons, and neutrinos. For higher magnetic fields, those pions and muons would suffer significant energy losses through synchrotron radiation before decaying (see e.g. \cite{MuckeProtheroe2001a}, \cite{Muckeetal2003} and \cite{ReynosoRomero2009}) The electrons and positrons
cool by producing photons, which can, in turn, interact
with protons, in turn creating more secondaries. At the same time, neutrinos
will escape, while neutrons, depending on the opacity,
can either escape, interact, or decay inside the source. Since the target material in the studied environment has low density, proton-proton interactions and bremsstrahlung can be neglected.
To treat this complicated system,
we used the kinetic equation approach as described
in \cite{mastikirk95}; henceforth MK95, and 
\cite{mastiprothkirk}; henceforth MPK05.
However, in the present treatment we have extended
the set of equations by including, in addition to the three
for protons, electrons/positrons, and photons, two more
equations, one
for neutrons and one for neutrinos. The target photons and all other particle distributions are again taken to be isotropic.

As in MPK05 we can write
the kinetic equations for a homogeneous region containing
relativistic hadrons and leptons in the compact form
 \eqb
{{\partial n_i}\over{\partial t}}+L_i +Q_i=0.
\label{kineqgen}
 \eqe
Here, index $i$ refers to protons (denoted by `p'),
electrons/positrons (`e'),
photons (`${\rm \gamma}$'), neutrons (`n') and neutrinos (`${\rm \nu}$').
The functions to be determined are the dimensionless differential number
densities of the five species, which can be produced/injected
in the source through the operators $Q_i$
and destroyed/escape through the operators $L_i$.
They are normalised from the ordinary differential number densities $\hat n_i$ in the following way:
 \eqb
n_i(\gamma_i,t){\rm d}\gamma_i = \sigma_{\rm T} R \hat n_i(E_i,t){\rm d}E_i \qquad {\rm with} \qquad \gamma_i={E_i \over {m_ic^2}}
\label{normhadronel}
 \eqe
where $R$ is the radius of the source and $\sigma_{\rm T}$ the Thomson cross section. The
photon and neutrino energy is normalised with respect to the electron rest mass.
Time is also normalised, to the light-crossing time
of the source $\hat{t}_{\rm cr}=R/c$, so $t=c \hat t/R$,  where $\hat t$ is the actual time.

The processes that we have included are
\begin{enumerate}
\item Proton-photon pair production,

\item Proton-photon pion production,

\item Proton synchrotron radiation,

\item Electron synchrotron radiation,

\item Synchrotron self absorption,

\item Electron inverse Compton scattering,

\item Photon-photon pair production,

\item Electron-positron annihilation,

\item Compton scattering of photons on cool pairs,

\item Triplet pair production.
\end{enumerate}

The modelling of the leptonic processes (4-9) has been discussed in MK95 and \cite{mastikirk97}.
Photopair production (1) has been extensively discussed and modelled in MPK05.
Modelling of proton synchrotron radiation (3) was done in a manner analogous to
electron synchrotron. Triplet pair production (10) was modelled according to
\citep{MMB86,masti91}.
However, the basic improvement of the present paper is for
photopion production (2). This process was modelled in MK95
with ${\rm \delta-}$function approximations and without taking neutrons and neutrinos into account
in the injection and energy loss terms. Here we revisit the process
by doing a modelling based on the results of SOPHIA \citep{SOPHIA2000}, which we discuss
in the next section. A similar modelling of the above processes was recently undertaken by \cite{VieyroRomero2012}, but in a strongly magnetised and ``dirtier'' environment, with a stronger emphasis on proton-proton interactions and a weaker emphasis on the precise modelling of photopion interactions, which uses the approximations discussed in \cite{Romero2010}.

By including the various relevant terms, the kinetic equations for each species become

- Protons

 \eqb
{{\partial n_{\rm p}} \over {\partial t}} + L^{\rm BH}_{\rm p} + L^{\rm photopion}_{\rm p}+ L^{\rm psyn}_{\rm p} + {n_{\rm p} \over t_{\rm p, esc}} = Q_{\rm p}^{\rm inj} + Q^{\rm photopion}_{\rm p} + Q^{\rm ndecay}_{\rm p};
\label{kineqpro}
 \eqe

- Electrons

 \eqb
{{\partial n_{\rm e}} \over {\partial t}} + L^{\rm syn}_{\rm e} + L^{\rm ics}_{\rm e} + L^{\rm ann}_{\rm e}  + L^{\rm tpp}_{\rm e} +
{n_{\rm e} \over t_{\rm e, esc}} = \nonumber \\
Q^{\rm ext}_{\rm e} + Q^{\rm BH}_{\rm e} + Q^{\rm \gamma \gamma}_{\rm e} + Q^{\rm photopion}_{\rm e} + Q^{\rm tpp}_{\rm e} + Q^{\rm ndecay}_{\rm e};
\label{kineqele}
 \eqe

- Photons

 \eqb
{{\partial n_{\rm \gamma}} \over {\partial t}} + {n_{\rm \gamma} \over {t_{\rm \gamma, esc}}} + L^{\rm \gamma \gamma}_{\rm \gamma} + L^{\rm ssa}_{\rm \gamma} = \nonumber \\
Q^{\rm syn}_{\rm \gamma} + Q^{\rm psyn}_{\rm \gamma} + Q^{\rm ics}_{\rm \gamma} + Q^{\rm ann}_{\rm \gamma} + Q^{\rm photopion}_{\rm \gamma};
\label{kineqphot}
 \eqe

- Neutrinos

 \eqb
{{\partial n_{\rm \nu}} \over {\partial t}} + {n_{\rm \nu} \over t_{\rm esc}} = Q^{\rm photopion}_{\rm \nu} + Q^{\rm ndecay}_{\rm \nu};
\label{kineqninos}
 \eqe

- Neutrons

 \eqb
{{\partial n_{\rm n}} \over {\partial t}} + L^{\rm photopion}_{\rm n} + L^{\rm ndecay}_{\rm n} + {n_{\rm n} \over t_{\rm esc}} = Q^{\rm photopion}_{\rm n}.
\label{kineqneut}
\eqe

The operators are labelled according to the processes that produce them, i.e. proton-photon
pair production (BH), proton-photon pion production (photopion), electron synchrotron (syn), proton synchrotron (psyn), synchrotron self absorption (ssa), inverse Compton scattering (ics), photon-photon pair production ($\gamma \gamma$), triplet pair production (tpp), neutron decay (ndecay) and electron-positron annihilation (ann), while (ext) refers to external injection.
$t_{\rm p, esc}$ and $t_{\rm e, esc}$ are the escape times for protons and electrons, respectively, {\rm while $t_{\rm esc}$ is the escape time for neutrally charged particles}.
Compton scattering of photons on cool pairs is approximated by multiplying the photon escape term, $t_{\rm \gamma,esc}$, by the factor

\eqb
{t_{\rm \gamma, esc}} = t_{\rm esc} \times (1+H(1-x)\tau_{\rm T}/3)^{-1}
\eqe
where $\tau_{\rm T}$ is the Thomson optical depth and $H(1-x)$ is the Heavyside function \citep{lightzdz87}, with $x={h \nu}/ {m_ec^2}$ the dimensionless photon energy.

\subsection{Photopion interactions: Loss and source terms}
\label{subsec:photopion}

Photopion interactions produce a distribution of neutrons that is similar in energy to the outgoing protons, since the two basic channels

 \eqb
(a)\qquad p + \gamma \longrightarrow p + \pi^0 \label{eq:pp}
 \eqe
 \eqb
(b)\qquad p + \gamma \longrightarrow n + \pi^+ \label{eq:pn}
 \eqe
are about equally probable. Neutrons can escape the emission region essentially in a crossing
time, but they are susceptible to two processes along the way. They can interact with ambient photons in much the same way as protons, leading to a mirror image of the previous channels

 \eqb
(c)\qquad  n + \gamma \longrightarrow n + \pi^0 \label{eq:nn}
 \eqe
 \eqb
(d)\qquad  n + \gamma \longrightarrow p + \pi^- \label{eq:np},
 \eqe
or they can decay into protons, with a mean lifetime of $\tau=881.5 \pm 1.5 s $ \citep{Nakamura2012}

 \eqb
(e)\qquad  n \longrightarrow p + e^- + \bar \nu_{\rm e} \label{eq:nd}.
 \eqe

In channels (a) - (d), the initial nucleons lose a portion of their energy that depends on the inelasticity of the interaction, $k_p$, but that is not a uniform process. SOPHIA simulation results show a distribution of nucleons after each interaction, with their maximum energies close to that of the initial nucleon and their minimum energies forming a tail that can extend over several orders of magnitude. Although these distributions have recently been approximated analytically \citep[see][]{Kelner08}, we made use of them in their original form, representing their effect with a coefficient $d(\gamma,\xi,\gamma')$, where $\gamma$ is the resulting nucleon's energy,  $\xi$ the target photon energy, and $\gamma'$ the initial nucleon's energy. Each interaction can be seen as removing the initial nucleon from its energy bin and creating the distribution of new ones, so that translates into a catastrophic loss term coupled with an injection term. The cross section $\sigma_{\rm N}(\gamma',\xi)^{\rm photopion}$ is related to the interaction time $\tau_{\rm N}(\gamma',\xi)^{\rm photopion}$ for a single nucleon in a monoenergetic photon field of density $1{\rm cm}^{-3} $, so that $\sigma_{\rm N}(\gamma',\xi)^{\rm photopion}=1/(\tau_{\rm N}(\gamma',\xi)^{\rm photopion}c)$, where N is the type of nucleon.

Therefore,
the contributions to the loss and injection terms in the proton and neutron kinetic equations from the above processes are (respectively)

-- Channels (a) and (b)

 \eqb
  Q_{\rm p}^{\rm \gamma p \rightarrow p \pi}(\gamma,t) = \sum_{\gamma',\xi} \sigma_{\rm p}(\gamma',\xi)^{\rm photopion} d(\gamma,\xi,\gamma') n_{\rm p}(\gamma',t)n_{\rm \gamma}(\xi,t) 
 \eqe

 \eqb
  L_{\rm p}^{\rm \gamma p \rightarrow p \pi}(\gamma,t) &=& -\sum_\xi \sigma_{\rm p}(\gamma',\xi)^{\rm photopion} n_{\rm p}(\gamma,t)n_{\rm \gamma}(\xi,t)
 \eqe

 \eqb L_{\rm p}^{\rm \gamma p \rightarrow n \pi}(\gamma,t) = -\sum_\xi \sigma_{\rm p}(\gamma',\xi)^{\rm photopion}  n_{\rm p}(\gamma,t)n_{\rm \gamma}(\xi,t)
 \eqe

 \eqb
 Q_{\rm n}^{\rm \gamma p \rightarrow n \pi}(\gamma,t) = \sum_{\gamma',\xi} \sigma_{\rm p}(\gamma',\xi)^{\rm photopion}  d(\gamma,\xi,\gamma') n_{\rm p}(\gamma',t)n_{\rm \gamma}(\xi,t);
 \eqe

-- Channels (c) and (d)

 \eqb
Q_{\rm n}^{\rm \gamma n \rightarrow n \pi}(\gamma,t) &=& \sum_{\gamma',\xi} \sigma_{\rm n}(\gamma',\xi)^{\rm photopion}  d(\gamma,\xi,\gamma') n_{\rm n}(\gamma',t)n_{\rm \gamma}(\xi,t) 
 \eqe

 \eqb
L_{\rm n}^{\rm \gamma n \rightarrow n \pi}(\gamma,t) &=& -\sum_\xi \sigma_{\rm n}(\gamma',\xi)^{\rm photopion} n_{\rm n}(\gamma,t)n_{\rm \gamma}(\xi,t)
 \eqe

 \eqb L_{\rm n}^{\rm \gamma n \rightarrow p \pi}(\gamma,t) = -\sum_\xi \sigma_{\rm n}(\gamma',\xi)^{\rm photopion} n_{\rm n}(\gamma,t)n_{\rm \gamma}(\xi,t)
 \eqe

 \eqb
Q_{\rm p}^{\rm \gamma n \rightarrow p \pi}(\gamma,t) = \sum_{\gamma',\xi} \sigma_{\rm n}(\gamma',\xi)^{\rm photopion}  d(\gamma,\xi,\gamma') n_{\rm n}(\gamma',t)n_{\rm \gamma}(\xi,t).
 \eqe

Neutron decay produces complementary loss and injection terms for the neutrons and protons, respectively:

 \eqb L_{\rm n}^{\rm n \rightarrow p}(\gamma,t) = -{n_{\rm n}(\gamma,t) \over {\gamma \tau}},
 \eqe

 \eqb Q_{\rm p}^{\rm n \rightarrow p}(\gamma,t) = {n_{\rm n}(\gamma,t) \over {\gamma \tau}}  \label{eq:nd2}.
 \eqe

However, since neutrons are produced predominantly with energies close to that of the
protons undergoing photopion interactions, their Lorentz factors are high enough to ensure that neutron decay occurs mostly outside the source.
We ignore the ensuing interactions of the products. 
Neutrinos are only produced in photopion interactions, and they freely escape the emission region, with an escape time equal to the crossing time of the source. In the code the neutrino flux is the sum of all flavours since we may assume that there will be complete mixing after propagation.
Since neutrinos escape without interaction, their spectra is proportional to their injection spectrum. We neglect neutrinos from neutron decay.

\section{Modelling photopion interactions}

\subsection{The SOPHIA code}
\label{subsec:SOPHIA}

At low interaction energies, photohadronic processes are dominated by the $\Delta(1232)$ resonance, so their cross section can be
 approximated by a $\delta$-function. However, such an approximation neglects processes that become dominant for higher energies 
and are still important in lower ones. To remedy this, the SOPHIA  
Monte-Carlo event generator was developed \citep[see][]{SOPHIA2000}, taking into account the following interaction processes: 
resonance production, direct pion production and diffractive and non-diffractive multipion production.

\subsection{Modeling SOPHIA results}

The photopion spectra are calculated indirectly from the results of this SOPHIA Monte Carlo code. Protons of three distinct energies ($\gamma_{\rm p}=10^{6.1}, \gamma_{\rm p}=10^{9.1}, \gamma_{\rm p}=10^{12}$, where the subscript ${\rm p}$ refers to protons) are allowed to interact with isotropically distributed target photons of energies ranging from $\xi=10^{-9.1}$ to $\xi=10^{0.1}$ in logarithmic steps of 0.1. This provides us with interaction rates and energy distributions of all secondary particles, including protons (since the original protons are treated as having sustained catastrophic losses, giving rise to a new distribution over a wide range of energies) for those three specific initial proton energies. Specifically, we obtain an energy loss term for protons, and injection terms for photons, electrons, protons, neutrons, and neutrinos. The last result from pions and muons, which are assumed to decay instantaneously.

The energy grids used in our program are equally spaced in the logarithm of $\gamma$, with a resolution of ten bins per decade for protons and electrons and five bins per decade for photons. Furthermore, the lowest grid point for both protons and electrons is $\gamma_{\rm min}=10^{0.1}$, while for photons it is taken to be $x_{\rm min}=b\gamma_{\rm min}^2$, since the softest photons are assumed to be produced by electron synchrotron radiation, $b = B/B_{\rm c}$ being the magnetic field in units of the critical field $B_{\rm c} = m_{\rm e}^2c^3/(e\hbar) = 4,414 \times 10^{13}$ G.

Each particle's injection term is calculated as the energy distribution term from SOPHIA over the interaction time for the initial proton and photon energies, times the number densities of those initial protons and photons, all normalised to our dimensionless units,
 \eqb
Q^{photopion}_i(\gamma,t)={d(\gamma,\xi,\gamma')n_{\rm N}(\gamma',t)n_{\rm \gamma}(\xi,t) \over {\tau_{\rm N\pi}(\gamma',\xi)}}   \eqe
where $d(\gamma,\xi,\gamma')$ is the energy distribution term, as described in sect. \ref{subsec:photopion}, $\tau_{N\pi}(\gamma',\xi)$ is the interaction time, N the type of nucleon, $i$ the particle type, and $\gamma$ is substituted by $x$ for photons and $E_\nu / m_ec^2$ for neutrinos.

Since the produced spectra depend on the product $\gamma_{\rm p}\cdot \xi$, 
for any proton of energy $\gamma_{\rm p}$ interacting with a photon of 
energy $\xi$ the effects will be identical to those of a proton of energy 
$\gamma_{\rm p}^*$ interacting with a photon of energy 
$\xi^*=\xi \gamma_{\rm p} / {\gamma_{\rm p}^*}$, the only difference 
being that the energies of the secondaries will be shifted by 
$\gamma_{\rm p}^* / \gamma_{\rm p}$. Therefore, we can use the three 
distinct proton energies 
($\gamma_{\rm p}=10^{6.1}, \gamma_{\rm p}=10^{9.1}, \gamma_{\rm p}=10^{12}$) 
to calculate the spectra for protons of all intermediate or lower energies 
with this scaling approach.  These three energies were chosen for technical 
reasons (in the present paper injected protons have $\gamma_p \ll 10^{12})$. 
It comes at an acceptably small loss of accuracy but at a gain in 
computing time and memory. 

When the initial particle is a neutron instead of a proton, we can use 
the same data but with switched labels; neutrons instead of protons, 
and all particles switched with their antiparticles. Comparisons with 
SOPHIA results from neutron-photon interactions show this method to be 
sufficiently accurate, since the error in energy of the switched distributions is no higher than 3.5\%.

\subsection{Energy losses}

To test the additions to the code, we performed both
single-interaction checks and full runs for various cases
of monoenergetic protons injected into black body radiation fields.
In Fig.~\ref{mono_dist} we show the distributions resulting
from interactions of monoenergetic protons of Lorentz factor
$\gamma_{\rm p}=10^{6.1}$ with photons with $\xi=10^{-2.8}$.
There, 95.6\% of the proton's energy is conserved in the secondaries,
whose contribution by particle type corresponds almost perfectly
to the ones in the SOPHIA data. If we run the same test with photons
of $\xi=10^{-3.8}$ and $\xi=10^{-0.2}$, corresponding to the lowest
and highest energy photons for which we can test the results directly,
the energy conserved in the secondaries is 90\% and 98\%, respectively.

\begin{figure}

\resizebox{\hsize}{!}{\includegraphics{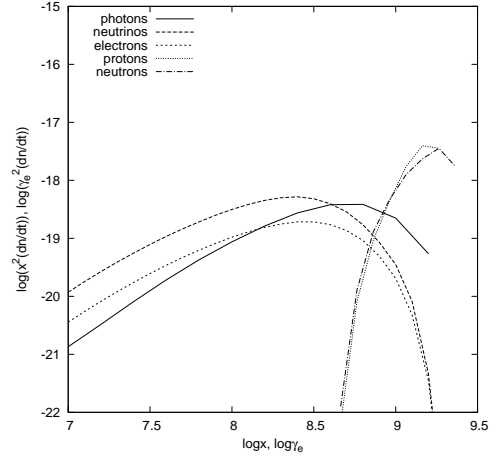}}

\caption{\small{Test distributions from the interaction of monoenergetic 
protons  ($\gamma_{\rm p}=10^{6.1}$) with photons ($\xi=10^{-2.8}$) 
as produced by the numerical code. Units in the y axis are arbitrary. 
These have been checked against the SOPHIA results and are in very 
good agreement.}}
 \label{mono_dist}
\end{figure}

%\begin{figure}
%
%\resizebox{\hsize}{!}{\includegraphics{test11024sophia.ps}}
%
%\caption{\small{Distributions from the same interaction from SOPHIA. Units in the y axis are arbitrary.}}
%\end{figure}

In Fig.~\ref{bb_loss} we show the energy losses
of protons in a black body radiation field (introduced merely for this test and not used elsewhere)
from both photopion interactions and pair production.
The relative contribution of each process for each proton
energy is as expected \citep{Stanevetal2000}.

\begin{figure}

\resizebox{\hsize}{!}{\includegraphics{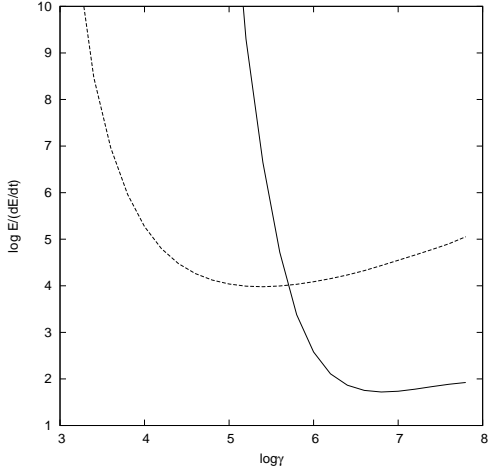}}

\caption{\small{Average energy loss times of protons interacting 
with ambient black body radiation of temperature 
$3\times 10^5K$ with $R=10^{15}{\rm cm}$ introduced here purely 
for the purpose of checking energy losses. Long-dashed line 
corresponds to pair production, while full line corresponds 
to photopion interaction}. Units in the y axis are dimensionless, 
normalised to the light crossing time.}
 \label{bb_loss}
\end{figure}

%\section{Extracting energy from the protons}

\section{Monoenergetic proton injection}

\subsection{Spectral signatures}

Adopting the standard picture of the one zone radiation
model, we can assume a spherical volume of radius $R$ and
a tangled magnetic field of strength $B$.
We consider
only the case where protons are injected in the system
and set the external injection of electrons equal to
zero.
The simplest way of treating proton injection in Eq. (\ref{kineqpro}) is to assume that
protons are injected with a ${\rm \delta-}$function at some energy $\gamma_{\rm 0}$. This
will faciliate some tests of the radiative signatures
of the different hadronic processes discussed in the
previous paragraph. Thus we write
\eqb
Q_{\rm p}^{\rm inj}(\gamma_{\rm p})=
Q_{\rm p,0}\delta(\gamma_{\rm p}-\gamma_{\rm 0})
\label{qpinject}
\eqe
where
$Q_{\rm p,0}$ is the proton normalisation, considered
as independent of time.
In the present treatment we find it more convenient, instead of
defining $Q_{\rm p,0}$, to equivalently define the proton compactness which is
given by the relation (as in MK95).
\eqb
\ell_{\rm p}= {m_{\rm p} \over {3m_{\rm e}}} \int d \gamma_{\rm p} (\gamma_{\rm p} -1) Q_p^{\rm inj}(\gamma_{\rm p}).
\label{lpdef}
\eqe
With the ${\rm \delta-}$function assumption, we can write
\eqb
Q_{\rm p,0}={{3m_{\rm e}}\over{m_{\rm p}}}{{\ell_{\rm p}}\over{(\gamma_{\rm 0}-1)}}.
\eqe
As was shown in MPK05, the actual injected proton luminosity
is related to $\ell_{\rm p}$ by the relation
\eqb
L_{\rm p}={{4\pi R m_{\rm p} c^3} \over{\sigma_{\rm T}}}\ell_{\rm p}.
\eqe

It will also prove useful to define, in a similar manner, the compactnesses
\eqb
\ell_i= {m_{\rm p} \over {3m_{\rm e}}} \int d \gamma_{\rm p} (\gamma_{\rm p} -1) L_{\rm p}^i(\gamma_{\rm p}).
\label{lidef}
\eqe
where $i$ refers to each process that creates a loss term for protons in Eq.~(\ref{kineqpro}). 

In addition to
$\gamma_{\rm 0}$, $\ell_{\rm p}$,
and $t_{\rm p,esc}$, one needs to specify
initial conditions for the five unknowns
to fully determine the system. Without loss of generality, we
can assume that at $t=0$, $n_{\rm p}(\gamma_{\rm p},0)=n_{\rm e}(\gamma_{\rm e},0)=n_\gamma(x,0)
=n_{\rm n}(\gamma_{\rm n},0)=n_{\rm \nu}(E_{\rm \nu},0)=0$. 
Then we can integrate the system forward in time.

Therefore, qualitatively speaking, one expects that
for $t>0$, protons
will start accumulating in the source. At the same time,
according to Eq.~(\ref{kineqpro})  protons will lose energy
by synchrotron and, possibly, by photopair and photopion production, while a fraction
will physically escape at a rate $t^{-1}_{\rm p,esc}$ from the source region.
In this case secondaries will be created making
Eqs.~(\ref{kineqele}) to (\ref{kineqneut}) relevant for the evolution of the system.

As a first example, we show the results obtained when
$R=3\times 10^{16}$~cm, $B=1~{\rm G}$, $\ell_{\rm p}=0.4$,
$t_{\rm p,esc}=t_{\rm cr}$, and $\gamma_0=2.5\times 10^6$. The magnetic field 
strength is such as would be expected at a distance of roughly 0.1pc 
from the base of the jet (\cite{Komissarov2007}, \cite{OSullivan2010}).
The parameters were chosen in such a way that the radiated photons will cause
minimal losses on the protons, therefore the proton steady-state
distribution is very close to the one derived from Eq.~(\ref{kineqpro})
when all loss terms are ignored, i.e.
\eqb
n_{\rm p}(\gamma_{\rm p})={{3m_{\rm e}}\over{m_{\rm p}}}{{{\ell_{\rm p}}{t_{\rm p,esc}}}\over{\gamma_{\rm 0}-1}}
\delta(\gamma_{\rm p}-\gamma_{\rm 0}).
\eqe
This simple form of the proton distribution function
faciliates the investigation of
some important points regarding the injected electron and
radiated photon spectrum. For this reason we do not 
consider the presence of an external photon distribution.
In this and the next test cases, the proton synchrotron 
photons serve as targets. 

\begin{figure}
\resizebox{\hsize}{!}{\includegraphics{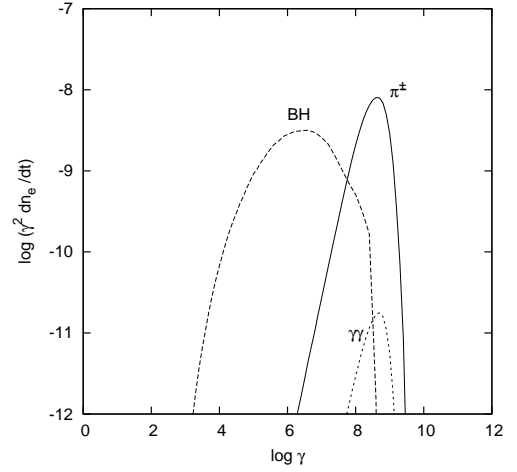}}
\caption{\small{Production rate of secondary
electrons for
$R=3\times 10^{16}$~cm, $B=1~{\rm G}$, $\ell_{\rm p}=0.4$,
$t_{\rm p,esc}=t_{\rm cr}$ and $\gamma_0=2.5\times 10^6$.
Full line depicts the injection resulting from charged
pion decay while long-dashed line is the photopair
(Bethe-Heitler) injection. Short-dashed line depicts
the pairs injected from $\gamma\gamma$ absorption,
which is negligible for the particular
set of the initial parameters
chosen.}}
\label{elecrate}
\end{figure}

At low compactnesses, as we are considering here,
electrons are injected into the system as secondaries mainly
through two channels
(i) from photopair and (ii) from photopion.
(At higher compactness photon-photon pair production
becomes important but we can neglect this for the time being.)
These two processes are in direct competition with each other
and their relative importance depends on such parameters as $\gamma_{\rm 0}$
and the soft photons which serve as targets.
Figure~\ref{elecrate}  plots the injection functions
of these two processes for the initial parameters given above.
We note that the two distributions have different characteristics.
The injection function of photopair electrons is broader and
has a peak at energies $\gamma_{\rm e}\simeq\gamma_{\rm p}$. The injection
function of photopion electrons, on the other hand, is flatter
and peaks at much higher energies, of the order of
$\gamma_{\rm e}=\eta_{\rm \pi e}\gamma_{\rm p}$,
with $\eta_{\rm \pi e}\simeq 150$.

In the case we are considering here, the photon spectrum
will show four distinctive features. Two of them are connected
to the synchrotron radiation of the injected electron populations
discussed above, while the two other features are connected to
proton synchrotron and $\pi^{\rm 0}-$decay. Putting all of them in ascending
order with frequency we have

\begin{enumerate}

\item {{\sl Proton synchrotron radiation:} Since the proton distribution
function is a ${\rm \delta-}$function at $\gamma_{\rm 0}$, the radiated photon
spectrum will have a peak at $x\simeq
{{m_{\rm e}}\over{m_{\rm p}}}b\gamma_{\rm 0}^2$ (where $b = B/B_{\rm c}$ , $B_{\rm c} = m_{\rm e}^2c^3/(e\hbar)$).}

\item{{\sl Synchrotron radiation from photopair electrons:}
As stated above, the electron injection function resulting
from photopair interactions is rather broad with a peak at
$\gamma_{\rm e}\simeq\gamma_{\rm p}$. Synchrotron cooling of electrons
and consequent radiation results in a photon spectrum with peak
at $x\simeq b\gamma_{\rm 0}^2$.}

\item{{\sl Synchrotron radiation from photopion electrons:}
In complete analogy to the photopair, the peak of this
distribution will be at $x\simeq b(\eta_{\rm \pi e}\gamma_{\rm 0})^2$.}

\item{{\sl$\gamma$-rays from $\pi^{\rm 0}-$decay:} A monoenergetic proton
distribution produces a well defined peak at
$x\simeq\eta_{\pi\gamma}\gamma_{\rm 0}$, with $\eta_{\pi\gamma}\simeq 350$.}
\end{enumerate}

\begin{figure}
\resizebox{\hsize}{!}{\includegraphics{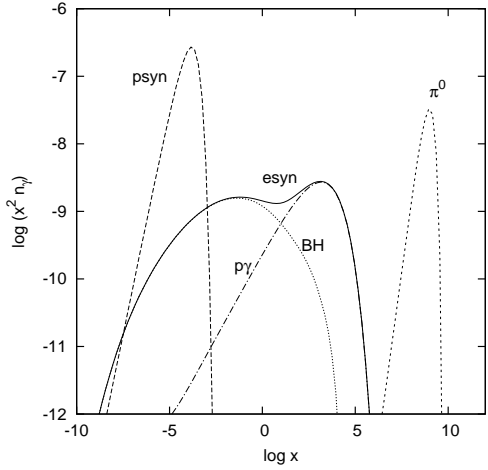}}
\caption{\small{Steady state MW spectrum of photons
resulting from monoenergetic proton injection with minimal cooling.
Parameters are as in the previous figure.
Full line is the synchrotron spectrum corresponding to the
electrons injected from photopair (dotted line) and photopion
(dot-dashed line) -- see previous figure. Long dashed line is
the proton synchrotron component while short dashed
line are the $\gamma-$rays resulting from $\pi^{\rm 0}$ decay.
}}
\label{photrate}
\end{figure}

It is interesting to note that the ratio of where the three first peaks occur
is fixed, i.e. $(m_{\rm e}/m_{\rm p}):1:\eta_{\rm \pi e}^2 = 5 \times 10^{-4} : 1 : 3 \times 10^{4}$, 
which implies that
the proton synchrotron peak will always be about eight orders of
magnitude below the synchrotron peak of photopion electrons.
Only the $\gamma-$ ray peak from neutral pion decay
is not connected to the other three -- since it does not involve
the magnetic field, but for all astrophysically relevant magnetic
field values, it should always be the highest energy peak.
These features can be seen in Fig.~\ref{photrate}
where the multi-wavelength (MW) spectrum corresponding to the parameters adopted above
is plotted. The inverse Compton scattering component
of the electrons is negligible.

\subsection{Increasing the injected proton energy: The transition 
in importance from photopair to photopion}

While protons will always radiate through
synchrotron radiation, it is a matter of the choice of the initial
parameters whether they will undergo substantial photopair and photopion
production. The rate of these processes reaches a maximum
when the energy of the collision between the
peak of the proton synchrotron
radiation and the relativistic proton is above
their respective thresholds. 
If we denote the proton synchrotron typical energy by  
$x_{\rm psyn}={m_{\rm e} \over m_{\rm p}}b \gamma_0^2 $, then if
\eqb
  x_{\rm psyn} \gamma_{\rm 0} < 2,
\label{thresholdpair}
\eqe
then obviously the proton-photon collisions cannot produce copious
photopairs because most of the collisions occur
below threshold for this process.
If
 \eqb
 2 < x_{\rm psyn} \gamma_{\rm 0} < {m_{\rm \pi} \over m_{\rm e}}(1+{m_{\rm \pi} \over {2m_{\rm p}}}),
 \eqe 
then the collisions between $x_{\rm psyn}$ and $\gamma_{\rm 0}$ 
are above the threshold for photopair but not for
photopion; note however the synchrotron radiation from 
the produced
photopairs can produce photopions, therefore we expect such a component 
even in this case, albeit at a low level. Finally, for 
 \eqb
 {m_{\rm \pi} \over m_{\rm e}}(1+{m_{\rm \pi} \over {2m_{\rm p}}}) < x_{\rm psyn}\gamma_{\rm 0}
 \eqe 
we expect that protons can  produce both pairs and pions
directly in collisions with proton synchrotron photons. 

The above relations help us define two characteristic proton
energies that are functions of the magnetic field only;
i.e.,
\eqb
\gamma_{\rm p,pair}=\left({{2}\over{b}}{{m_{\rm p}}\over{m_{\rm e}}}\right)^{1/3}
\label{gpair}
\eqe
and
\eqb
\gamma_{\rm p,pion}=
 \left({1 \over b}{m_{\rm p} m_{\rm \pi} \over m_{\rm e}^2}
\left(1+{m_{\rm p} \over {2m_{\rm \pi}}}\right)\right)^{1/3}. 
\label{gpion}
\eqe
Depending on the proton injected energy
$\gamma_0$ with respect
to $\gamma_{\rm p,pair}$ and $\gamma_{\rm p,pion}$, the protons 
can produce either photopairs or both photopairs and photopions. 
We emphasise, however, that the above characteristic proton
energies are only indicative. Since we are using the full emissivities, 
rather than delta-function approximations for the energies of 
secondary particles and photons,
it is possible to have photopairs even if Eq. (\ref{thresholdpair})
does not strictly hold, as relativistic protons can always pair produce
with photons at the tail of the synchrotron distribution.

Figure~\ref{relativeinjection}
shows the MW photon spectrum
in the case of monoenergetic proton injection with
$\gamma_{\rm 0}=3\times 10^5$,~$3\times 10^6$,
and $3\times 10^7$,
while all the other parameters have been kept constant
to the values given in the previous paragraph.
The run for $\gamma_{\rm 0}=3\times 10^5$ marginally satisfies the photopair condition 
and not the photopion one. Thus the photon spectrum consists
of the proton synchrotron peak and the photopair synchrotron
emission.
The run for $\gamma_{\rm 0}=3\times 10^6$ satisfies both conditions,
and the photon spectrum shows all four
 features, as discussed in the previous section.
Finally, the run for $\gamma_{\rm 0}=3\times 10^7$ has a much stronger
photopion than photopair component, and as a result the synchrotron
signature of the latter lies below that of the former, resulting 
in only one peak for the synchrotron spectrum,
i.e. the photopion one.

To systematise the above results, we have plotted
the injected compactness of each individual proton process {\rm (as defined in 
Eq. (\ref{lpdef}), but for $Q_{\rm p}^{\rm inj}(\gamma)$ corresponding to 
each particular process, rather than to total proton injection) } 
in Fig.~\ref{injectg0}
as a function of $\gamma_{\rm 0}$, while all other parameters
have been kept constant. For the injected photopion
compactness we considered the sum of $\gamma-$rays
and electrons resulting from neutral and charged pion decay,
respectively, since both contribute to the photon spectrum.
We note that, while
the proton synchrotron term varies linearly with the injected
energy, the photopair and photopion terms show more complex
behaviour. Both increase initially reaching a maximum,
and subsequently decrease. As discussed above, photopair
is more important for lower injected
proton energies, but as the injection energy
increases, photopion becomes dominant -- for the
assumed magnetic field, $\gamma_{\rm p,pair}=5\times 10^5$,
while $\gamma_{\rm p,pion}=5\times 10^6$.
We also note that proton synchrotron
is more important as a loss/injection process than the
other two for all energies,
but this depends on the value of the
injected proton compactness. As we show in the next
section this trend changes as the proton compactness increases.

\begin{figure}
\resizebox{\hsize}{!}{\includegraphics[angle=270]{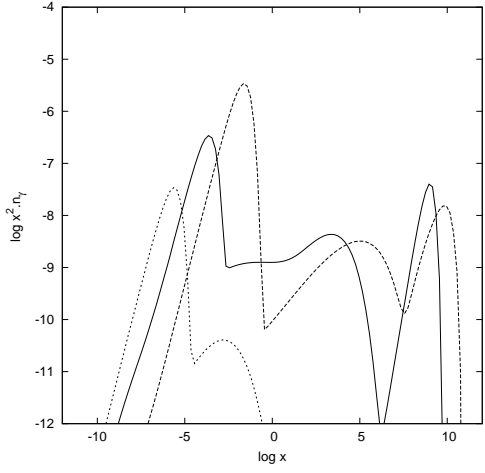}}
\caption{\small{Steady state MW spectrum of photons
resulting from a monoenergetic proton injection with
$\gamma_{\rm 0}=3\times 10^5$ (dotted line),~$3\times 10^6$ (full line),
and $3\times 10^7$ (dashed line).
The other parameters are
$R=3\times 10^{16}$~cm, $B=1~{\rm G}$, $\ell_{\rm p}=0.4$, and
$t_{\rm p,esc}=t_{\rm cr}$.
}}
\label{relativeinjection}
\end{figure}

\begin{figure}
\resizebox{\hsize}{!}{\includegraphics{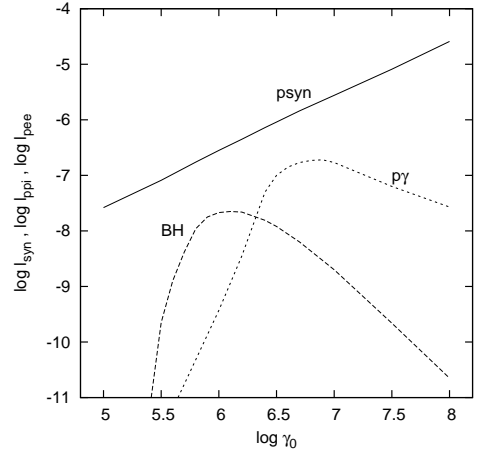}}
\caption{\small
{The compactnesses of proton processes for a ${\rm \delta-}$function
proton injection as a function of the energy
$\gamma_{\rm 0}$. Full line is for proton synchrotron,
dashed line for photopair, and dotted line for photopion.
For this process
 the sum of photon and electron injection resulting from
neutral and charged pion decay is plotted. The other parameters
are $R=3\times 10^{16}$~cm, $B=1~{\rm G}$, $\ell_{\rm p}=0.4$, and
$t_{\rm p,esc}=t_{\rm cr}$.} }
\label{injectg0}
\end{figure}

\subsection{Increasing the injected proton compactness: From linear
to non-linear proton cooling}

We turn next to investigating the effects that the injected proton
compactness has on the photon spectra. In the case of proton-only
injection like the one we are considering here, there are, up to a degree,
profound analogies to the synchrotron - SSC relationship of a leptonic
system. There the electrons radiate synchrotron photons, and
consequently upscatter them through
inverse Compton scattering interactions.
As long as the magnetic energy density dominates the synchrotron
photon density, the system can be considered in the linear
regime. This situation changes when the synchrotron photon density
dominates and the system becomes non-linear leading to the well-known
Compton catastrophe.

One can consider something analogous for the system we are
considering at present -- see also \S2 in \cite{petromasti12}.
In  hadronic systems and for
 the example we have examined above one can argue that the
system is also in the linear regime, because protons radiate by
synchrotron and the thus-radiated photons are used as targets
for photopair and photopion production. As can be seen from
Fig.~\ref{injectg0}, the synchrotron luminosity dominates,
which means that the cooling, however small, is regulated
by this process. 
Figure~\ref{injectg0} can also be viewed in terms of the energy-loss 
time scales for each process. Although the compactnesses depend 
on the proton density, the system's very low efficiency (as 
discussed in the next section) means that, for a given proton 
injection, that density will be practically constant. Then, 
for each proton injection 

\eqb
Q_{\rm p}^{\rm inj}(\gamma_{\rm p}) = {n_{\rm p} \over {t_{\rm p,esc}}},
\eqe
the loss terms will be 
\eqb
 L_{\rm p}^i(\gamma_{\rm \rm p}) = {{n_{\rm p}} \over {t_i}}
\eqe
where $i$ can be `syn', `BH' or photopion (`p$\gamma$'). Substituting in Eqs.~(\ref{lpdef}) and ~(\ref{lidef}), we find
\eqb
{t_i \over t_{\rm p,esc}} = {\ell_{\rm p} \over \ell_i}.
\eqe

Therefore the question that becomes relevant
is what happens to the system if
the proton injection luminosity is increased
further while the magnetic field value is kept
constant. This would essentially mean that the photon
density of the system increases over the magnetic
one, and as a result the photopair
and photopion losses/injection increase more than
the proton synchrotron ones. This occurs because, while
the synchrotron luminosity depends on the proton
density, the photopair and photopion luminosities depend
on both the proton and photon density. Since the photon density
depends on the proton density, we conclude that
the above processes depend quadratically on the proton density
provided that both use the proton synchrotron photons as 
targets, which holds whenever $\gamma_{\rm 0} > \gamma_{\rm p,pion}$
(c.f. Eq.~\ref{gpion}).

The above can be seen in
Fig.~\ref{deltalp}, which depicts
the steady state MW spectrum in the case where the injection
proton compactness takes the values $\ell_{\rm p}$=0.4,~1.3,~4,~13, and 40
(bottom to top). One notices that, as the injection
compactness increases,
the synchrotron component increases linearly while the photopair
and photopion increase quadratically. However, for $\ell_{\rm p}\simeq$40, 
the system undergoes a transition, and
the photon luminosity goes up by a factor of $~10^4$. 
These types
of transitions are a sign that the system becomes {\sl supercritical}
and are caused by various feedback mechanisms
-- like the pair production - synchrotron loop \citep{kirkmasti92}
and the automatic photon quenching one \citep{stawarzkirk, petromasti11} --
which cause a very fast, non-linear proton cooling with simultaneous
exponential increase in the secondaries. 
The system then can either reach a steady state, as in the
example shown above, or show limit cycles \citep{sternsvensson, 
mastiprothkirk,petromasti12}.
A systematic search of their effects are beyond the scope of
the present work, but we will investigate their properties 
fully in a forthcoming paper; see, however, \cite{dimis12}
for some preliminary results.
For the present, it suffices to say
that once the behaviour shown in Fig.~\ref{deltalp} is typical,
i.e. for fixed initial parameters, there is always 
a value of $\ell_{\rm p}$ above which the system becomes supercritical.
We define this value as
$\ell_{\rm p,cr}(R,B,\gamma_{\rm max})$ and we refer to it in \S 6.
A final remark concerning the above figure is that
the ratio between the photopair and photopion injection rates,
which has remained constant as long as the system was in the
subcritical regime, starts changing as the system becones supercritical. 
This is because the spectrum of the target photons changes causing
the respective rates to change accordingly.

\begin{figure}
\resizebox{\hsize}{!}{\includegraphics[angle=270]{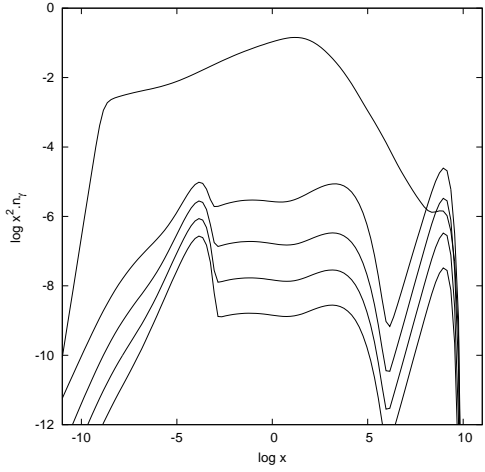}}
\caption{\small
{Steady state MW photon spectra for ${\rm \delta-}$function
proton injection at energy
$\gamma_{\rm 0}=2.5\times 10^6$ and different injection compactnesses
$\ell_{\rm p}=$0.4,~1.3,~4,~13, and 40 (bottom to top).
The other parameters are
$R=3\times 10^{16}$~cm, $B=1~{\rm G}$, $t_{\rm p,esc}=t_{\rm cr}$. }}
\label{deltalp}
\end{figure}

Figure~\ref{injectlp} shows the injected luminosity from each proton process.
We note that synchrotron remains linear throughout, while photopair
and photopion are initially quadratic before they become highly non-linear
once the proton density enters the supercritical regime.

\begin{figure}
\resizebox{\hsize}{!}{\includegraphics{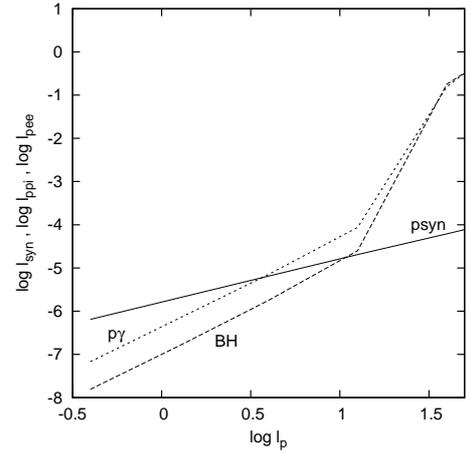}}
\caption{\small
{The compactnesses of proton processes for a ${\rm \delta-}$function
proton injection as a function of the proton compactness $\ell_{\rm p}$.
Full line is for proton synchrotron,
dashed line for photopair, dotted line for photopion
-- for this last process
 the sum of photon and electron injection resulting from
neutral and charged pion decay is plotted. The other parameters
are as in the previous figure.} }
\label{injectlp}
\end{figure}

An interesting result of the above analysis is the following.
In cases where $\gamma_{\rm p,pair} < \gamma_0 < \gamma_{\rm p,pion}$,
then, according to the discussion in the previous subsection,
protons can directly produce photopairs on proton synchrotron
photons but not photopions. The latter, however, can be still
created on the synchrotron radiation of the produced photopairs.
Thus, any variation in the injection rate will cause the
proton synchrotron photons to vary linearly, the radiation
from photopairs quadratically (as already explained above),
and the radiation fron photopions cubically, since their rate
depends on protons and photopairs. 
We return to this point in \S6.

\section{Power-law proton injection}

We next examine the case where the injected protons form a power law,
i.e. the injection function is

\eqb
Q_{\rm p}^{\rm inj}(\gamma)=
Q_{\rm p,0}\gamma^{-s}H(\gamma-\gamma_{\rm p,min})H(\gamma_{\rm p,max}-\gamma)
\label{qpinjectpl}
\eqe
where $\gamma_{\rm p,min}$ and $\gamma_{\rm p,max}$ are the lower 
and upper cut-offs of the proton distribution,
respectively, $Q_{\rm p,0}$ is the proton normalisation (considered, for the time being,
as independent of time), and $H(x)$ is the Heaviside function.
From Eq.~\ref{qpinjectpl} one can define a proton compactness in
complete analogy to the monoenergetic case  -- c.f. Eq.~\ref{lpdef}.

Figure \ref{all_gmax6} depicts the spectra of photons,
neutrons, and neutrinos of all flavours 
emerging from a source with parameters
$\gamma_{\rm p,min} = 10$, $\gamma_{\rm p,max} = 10^6$, $R=3 \times 10^{16}$ cm,
$B=1$~G,
$s = 2$, $\ell_{\rm p}=30$, and $t_{\rm p,esc}=t_{\rm cr}$.
For comparison reasons we have also plotted
the injected proton spectrum. The
neutrino spectrum peaks at high energies and has a distribution
that resembles that of $\gamma-$rays resulting from $\pi^{\rm 0}-$decay.
The neutrons, on the other hand, are more sharply peaked
with a maximum in their distribution that is very close to
$\gamma_{\rm p,max}$.
Another interesting point is that of efficiency,
i.e. of the fraction of the luminosity that goes into radiation, neutrons, and neutrinos
with respect to the total power injected into protons, which is quite low,
 at least for the parameters of the present example. Thus photons take
about $10^{-7}$ of the available luminosity in protons, while a comparable amount
goes to neutrons and neutrinos. We mention that for these low
compactnesses, neutrons escape the source practically unattenuated
and deposit their energy in the surrounding region once they are
transformed back to protons \citep{kirkmast89,GiovanoniKazanas}.
Their resulting radiation will peak at UHE energies from $\pi^0$ decay and at 
VHE energies from synchrotron cooling of produced protons \citep{MastiProth90} .

In Fig. \ref{MW_gmax_all}, which is analogous to Fig. \ref{relativeinjection}
for the monoenergetic injection case, we treat $\gamma_{\rm p,max}$ as a free parameter,
taking values
from $10^{5.5}$ to $10^8$ in increments of 0.5 in logarithm.
For $\gamma_{\rm p,max}=10^{5.5}$ the photopair threshold is just satisfied in 
collisions between proton synchrotron
photons and protons. Even so, there is a low-lumimosity photopion
component created from collisions between
synchrotron photons from photopair electrons and protons. Generally, in
close
analogy to the monoenergetic case, we find that
as $\gamma_{\rm p,max}$ increases, photopion dominates the photopair; however,
at very high energies of proton injection the spectrum starts saturating
as photon-photon absorption becomes dominant.

\begin{figure}
\resizebox{\hsize}{!}{\includegraphics{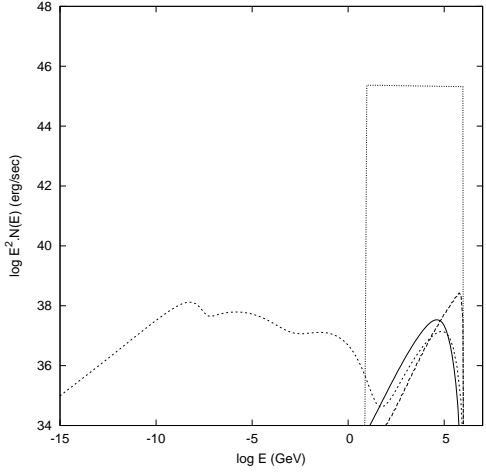}}
\caption{\small{Spectra of photons (small dashed line), neutrons (long
dashed line) and neutrinos (full line) escaping from the source for
initial parameters:
$s = 2$, $\ell_{\rm p}=30$, $t_{\rm p,esc}=t_{\rm cr}$,
$\gamma_{\rm p,min} = 10$, $\gamma_{\rm p,max} = 10^6$, $R=3 \times 10^{16}$ cm, and
$B=1$~G.
For comparison we have drawn the proton injected spectrum with a dotted line.
}}
\label{all_gmax6}
\end{figure}

\begin{figure}
\resizebox{\hsize}{!}{\includegraphics{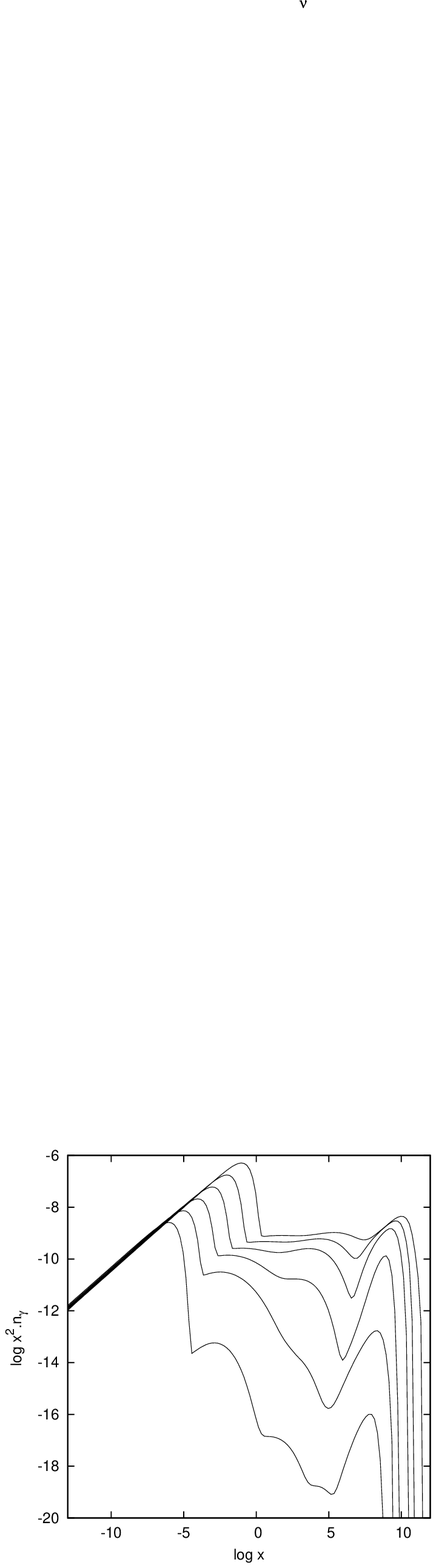}}
\caption{\small{Steady-state spectra of photons
resulting from a power-law proton injection
$s = 2$, $\ell_{\rm p}=0.3$, $t_{\rm p,esc}=t_{\rm cr}$,
$\gamma_{\rm p,min} = 10$, and  $\gamma_{\rm p,max} = 10^{5.5}-10^8$
with increments of 0.5 in logarithm (bottom to top).
The other parameters are $R=3 \times 10^{16}$ cm and
$B=1$ G.
}}
\label{MW_gmax_all}
\end{figure}

In Fig. \ref{pl6} we plot the steady state spectra as a function of the 
initial proton luminosity. It is analogous to Fig.~\ref{deltalp} and 
one notices again the two basic
features discussed for that figure, i.e. the quadratic nature of the photopair
and photopion processes as opposed to the linear behaviour of the proton synchrotron
radiation and the non-linear transition of the system
to high luminosities once the protons
have reached a certain critical density.

Figure \ref{powerlaws8} treats the proton injection slope
$s$ as a free parameter. Since in each case the injected proton 
luminosity is the same, injection of
harder power laws mean that more of the luminosity is concentrated in higher 
energies, resulting in flatter spectra for the proton synchrotron
component and also in an increase in the luminosity going to the photopair
and photopion components. The latter can
be seen more clearly in Fig. \ref{luminosities8}, where the resulting luminosities
are plotted against $s$. Because we adopted a high value of
the upper proton cutoff ($\gamma_{\rm p,max}=10^8$), photopion dominates photopair production
for flat injection spectra, while the two processes become comparable for steeper spectra.

The neutrinos emitted from those same proton power laws at steady state
are plotted in Fig. \ref{powerlaws_nt8}. Close to their upper cutoff they resemble
power laws, which are harder than the proton ones by a factor of about $(s -0.5)/2.5$.

\begin{figure}
\resizebox{\hsize}{!}{\includegraphics{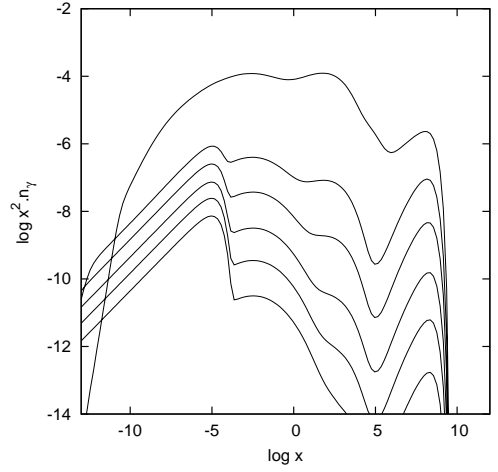}}
\caption{\small{MW steady-state spectra of photons
resulting from a power-law proton injection with
$\gamma_{\rm p,min} = 10$, $\gamma_{\rm p,max} = 10^6$,
$s = 2$, $t_{\rm p,esc}=t_{\rm cr}$,
and $\ell_{\rm p}$ taking the values of 0.3 to 100
with logarithmic increments of 0.5 (bottom to top).
The other parameters are $R=3 \times 10^{16}$ cm and
$B=1$ G.
}}
\label{pl6}
\end{figure}

\begin{figure}
\resizebox{\hsize}{!}{\includegraphics{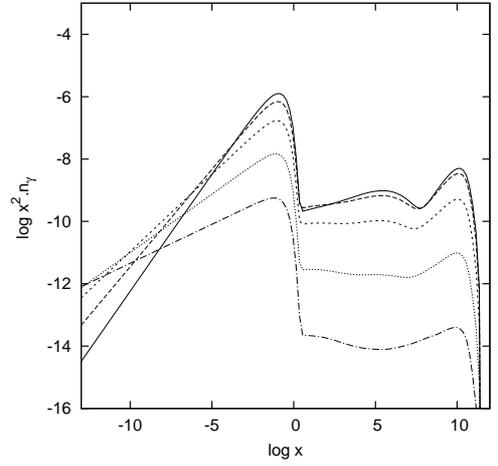}}
\caption{\small{MW steady-state spectra of photons
resulting from variable power-law proton injection with
$\gamma_{\rm p,min} = 10$, $\gamma_{\rm p,max} = 10^8$,
$\ell_{\rm p}=0.1$, $t_{\rm p,esc}=t_{\rm cr}$, $R=3 \times 10^{16}$ cm, $B=1$ G,
and $s$ taking the values of 1.5 to 2.5 with increments of 0.25 (top to bottom).
}}
\label{powerlaws8}
\end{figure}

\begin{figure}
\resizebox{\hsize}{!}{\includegraphics{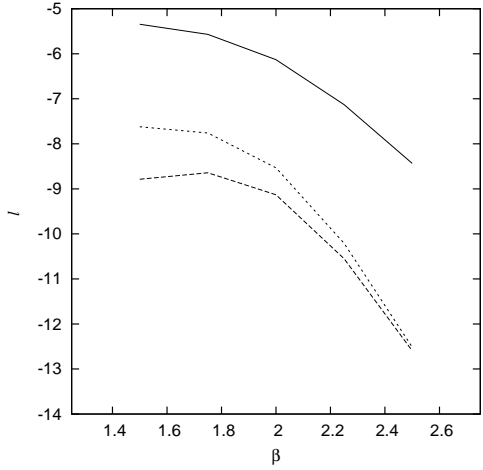}}
\caption{\small{Steady-state luminosity of photons produced by
proton synchrotron (full line), electron synchrotron from pair production
(long dashed lines), and electron synchrotron from photopion interactions (short dashed line),
plotted against $s$. All other initial parameters are as in the previous figure.
}}
\label{luminosities8}
\end{figure}

\begin{figure}
\resizebox{\hsize}{!}{\includegraphics{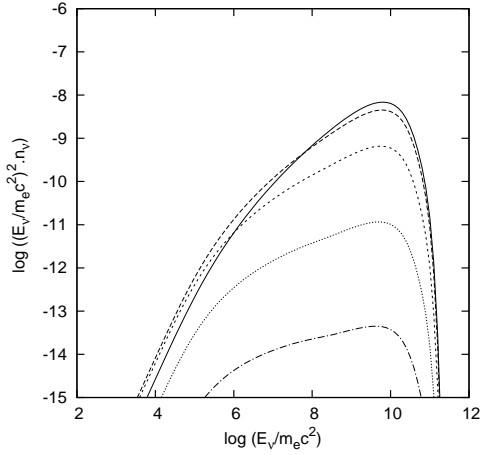}}
\caption{\small{MW steady-state spectra of (electron and muon) neutrinos
resulting from variable power-law proton injection as in Fig. \ref{powerlaws8}.
}}
\label{powerlaws_nt8}
\end{figure}

\section{Time variability}

The numerical scheme introduced through Eqs.
(\ref{kineqpro}) to (\ref{kineqneut})
can simulate time variability by treating one or more
of the free parameters as time-dependent.
Here we adopt the standard procedure followed in such cases;
i.e., we use initially constant injection parameters until
the system reaches a steady state and then we introduce
a perturbation in one of the parameters and record the
produced photon spectrum at each instant.

As an indicative example, we chose first as initial parameters those
that led to the steady state
obtained in Fig.~\ref{photrate} -- see also
Fig.~\ref{deltalp}, bottom curve.
We chose the particular example because, for this case, each hadronic component
leaves a clear signature on the MW spectrum, and thus it will be straightforward
to examine the time response of each component.
We also introduce a time variation to
the proton injection parameter $Q_{\rm p}^{\rm inj}$ in the form of a Lorentzian
profile
\eqb
f_{\rm L}(t;t_0,w,n)=1+(n-1){{w^2}\over{4(t-t_{\rm 0})^2+w^2}}.
\eqe
The above quantity reaches a maximum for $t=t_{\rm 0}$, i.e.
$f_{\rm L}(t_{\rm 0};t_{\rm 0},w,n)=n$, assuming $n>1$.
The quantity $w$ is the full width at half maximum
since $f_{\rm L}(t_{\rm 0}\pm w;t_{\rm 0},w,n)=(n+1)/2$. For the present
application the parameters used were $n=3.16$,
$t_{\rm 0}=100t_{\rm cr}$, and $w=10t_{\rm cr}$.

\begin{figure}
\resizebox{\hsize}{!}{\includegraphics[angle=270]{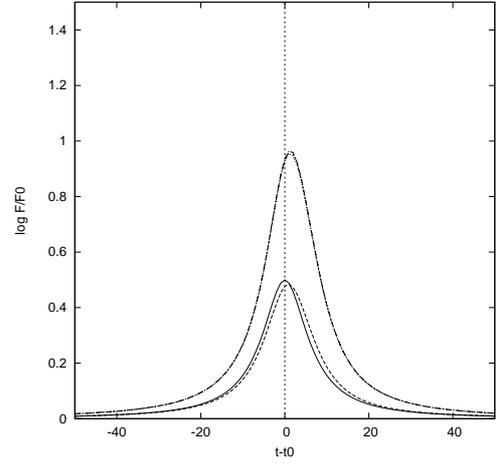}}
\caption{\small
{Energy centered lightcurves for photon energies 
$x_{\rm 1}=10^{-3.85}$ (dashed line),
$x_{\rm 2}=10^{-1.25}$ (dot-dashed line), and $x_{\rm 3}=10^{3.15}$ (short dashed line)
resulting from
a Lorentzian variation of
the proton injection parameter $Q_{\rm p}^{\rm inj}$ with $n=3.2$ and $w=10t_{\rm cr}$,
depicted here with full line.
This variation was introduced on a pre-existing steady-state
obtained with the parameters of figure~\ref{photrate} (or, equivalently,
fig.~\ref{deltalp}, bottom curve.) For this example lines $x_{\rm 2}$ and $x_{\rm 3}$
practically coincide.}}
\label{time1}
\end{figure}

Figure~\ref{time1} shows the resulting energy-centred (i.e. with the horizontal 
axis centred on the time when the proton energy injection becomes maximum)
photon lightcurves obtained for energies (expressed in logarithm)
$x_{\rm 1}=-3.84$, $x_{\rm 2}=-1.24$, and $x_{\rm 3}=3.15$. 
According to the discussion made following Fig.~\ref{photrate},
for the initial conditions used,
proton synchrotron contributes mainly to
$x_{\rm 1}$, while synchrotron from photopairs and
electrons from photopions to $x_{\rm 2}$, and electron synchrotron to $x_{\rm 3}$.
It becomes apparent that the lightcurve at energy $x_{\rm 1}$, to which
proton
synchrotron is contributing, follows the variation
in the proton injection very closely. The other energies have a clear
quadratic dependence on  proton variation, in accordance to the
discussion given in \S 4.1.

Figure~\ref{time2} repeats the calculations, but now 
we take a proton energy injection value that satisfies
the relation
 $\gamma_{\rm p,pair} < \gamma_{\rm 0} < \gamma_{\rm p,pion}$.
The bottom full line curve that corresponds to 
proton synchrotron varies linearly, while 
the middle and top ones, which are at the photopair and 
photopion electron characteristic
synchrotron energies, respectively, vary quadratically
and (almost) cubically. The reason behind this 
has already been explained in \S 4.1.

The above results are relevant when the system is in
the subcritical regime. The situation changes drastically when
the system is entering supercriticality. For example, 
Fig.~\ref{time3} shows the energy-centred lightcurves 
obtained from such a case. 
Comparing the produced lightcurves with those 
of the two previous figures, we find some interesting
differences. (i) 
Even though the proton 
compactness has been increased by a factor of 3,
the photon luminosity increases by a factor of $\sim$100
in all bands,
demonstrating the high non-linearity of the system.
(ii) The lightcurves peak about 10$t_{\rm cr}$
after the peak of the perturbation, while in the 
previous examples the time lags were shorter.
(iii)  The relation between the three components
discussed above is destroyed by the onset of 
the intense electromagnetic cascade caused by the
supercriticality. 

To further show the complex behaviour of the system 
when it becomes supercritical, we plot in Fig.~{\ref{time4}}
a flare caused by the sum of two identical Lorentzians which are
with peaks shifted by 30$t_{\rm cr}$. The ensuing
flare is asymmetrical with the second one producing much 
higher luminosity than the first one.

Therefore,
it becomes evident from the above that a hadronic system exhibits
far more complex behaviour than a leptonic one. The exact
relationship between its main radiating components (proton synchrotron,
photopair, and photopion) depends on both the injected proton energy
and the proton luminosity and needs further investigation.
Nevertheless, we present below, for the sake of completeness, a summary
of our previous results:

\begin{enumerate} 
\item{{\sl Subcritical regime:} For proton injection 
$\ell_{\rm p} < \ell_{\rm p,cr}(R,B,\gamma_{\rm max})$
the system behaves linearly; i.e., variations in proton injection
cause linear variations in the proton synchrotron component.
The photopair component, however, always varies quadratically,
while the photopion one varies either cubically (for 
$\gamma_{\rm p,max}\le \gamma_{\rm p,pion}$) or quadratically 
($\gamma_{\rm p,max} > \gamma_{\rm p,pion}$). In either case, the efficiency
of the system is rather low.}

\item{{\sl Supercritical regime:} For  
$\ell_{\rm p} > \ell_{\rm p,cr}(R,B,\gamma_{\rm max})$
the system undergoes an abrupt phase transition. At the onset
of it, even small perturbations in the proton injection
can cause high amplitude variations in the secondaries.
As in all dynamical systems, its exact behaviour
depends sensitively and in a non-linear way
on the initial conditions and the parameters
of the perturbation. On the other hand, the relation
between the various components lose the mostly quadratic
dependence found in the subcritical regime owing to the
electromagnetic cascading, which contributes to all bands. Finally, the efficiency
of the system can be quite high.}
\end{enumerate}

\begin{figure}
\resizebox{\hsize}{!}{\includegraphics[angle=270]{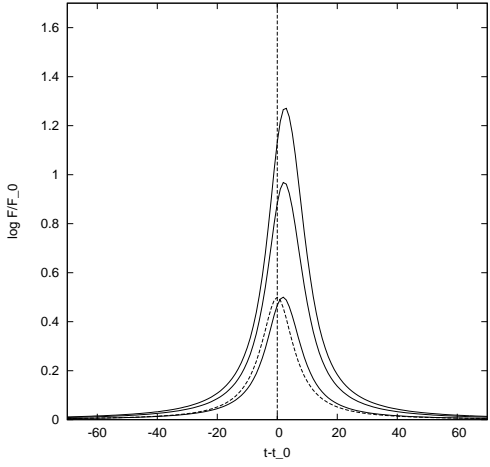}}
\caption{\small
{A flare produced in the case with 
$\gamma_{\rm p,pair} < \gamma_{\rm p,max} < \gamma_{\rm p,pion}$. Full lines
are at the characteristic energies for proton 
(bottom), photopair (middle) and photopion (top) synchrotron emission.}
}
\label{time2}
\end{figure}

\begin{figure}
\resizebox{\hsize}{!}{\includegraphics[angle=270]{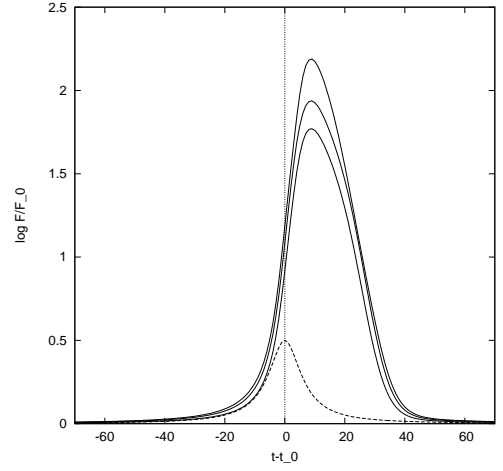}}
\caption{\small
{A flare produced when the system is supercritical. The parameters for the
Lorentzian are as in 
Fig.~\ref{time1}. The full lines are contributions from the various processes
with the same order as in Fig.~\ref{time2}}
}
\label{time3}
\end{figure}

\begin{figure}
\resizebox{\hsize}{!}{\includegraphics[angle=270]{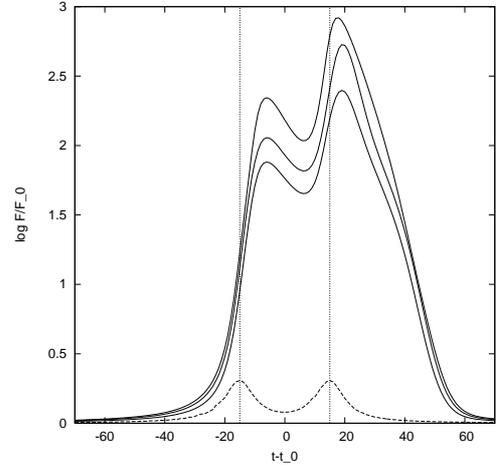}}
\caption{\small
{A flare produced when the system is supercritical by the sum of two Lorentzians
of the same amplitude. Here $n=2$ while the two peaks occur at
$t_{\rm 1}=85t_{\rm cr}$ and $t_{\rm 2}=115t_{\rm cr}$ respectively. For comparison
with the previous Figure, we have chosen $t_{\rm 0}=100t_{\rm cr}$.} 
}
\label{time4}
\end{figure}

\section{Discussion/Summary}

Hadronic models have been used extensively for fitting
the $\gamma-$ray emission of blazars. 
According to their basic premises, $\gamma-$rays can be
produced either directly via proton synchrotron radiation
or via the radiation of the secondaries resulting
from photopion collisions.
In the present paper which, in many respects, is an extension of 
earlier work by MK95 and 
MPK05, we have approached the model 
by writing and solving five coupled partial differential equations,
one for protons and the rest for the four stable species which
result from proton-photon interactions, namely electrons, photons,
neutrons and neutrinos, thus our approach is self-consistent
in the sense that it conserves energy.  
However, the major improvement of the
present work over MK95 and MPK05 was its 
modelling the spectra of 
secondaries from photopion interactions (\S 3) . For this we
have made systematic use of the SOPHIA results \citep{SOPHIA2000}
which were suitably parametrised in a way 
so the derived distribution functions of secondaries can serve 
as source functions for their corresponding 
equations. 
We then modelled the proton losses caused by these interactions so as
to maintain a detailed balance between the energy lost by protons
and that gained by the secondaries.
This  approach enables us to address topics such as the stability of the system,
the efficiency of proton luminosity conversion into radiation and
the relation between the underlying proton distribution with the
emitted photon and neutrino spectra.

Assuming a hadronic model with  a
monoenergetic proton injection 
and without any primary electron
injection (\S 4), we have showed that
for low proton energies the only important process, apart from
proton synchrotron radiation, is photopair (Bethe-Heitler) production
photopion operates, albeit at relatively low rates.
The resulting photon spectrum shows two features,
both of synchrotron origin, one from protons and one from
the produced Bethe-Heitler pairs.
For even higher proton injection energies, photopion becomes relevant
and the photon spectrum shows two more features, one due
to the synchrotron radiation from electrons and positrons
produced through charged pion decay and one
due to $\gamma-$rays from $\pi^{\rm 0}-$decay
(see Figs.~\ref{photrate} and \ref{relativeinjection}).
The relative importance of photopair and photopion depends critically
on the injected proton energy: for low proton energies photopair dominates,
while the situation is reversed for high energies (Fig.~\ref{injectg0}).

The above simple picture holds when the proton injected
luminosity (or compactness) is low and proton cooling minimal,
which means that the proton distribution function
does not deviate significantly from that at injection.
In this case the
proton synchrotron luminosity is dominant while
the other two processes contribute a smaller fraction to
the total radiated luminosity.
However, as the proton compactness increases, the photopair and
photopion losses increase quadratically.
 -- note, however, that depending on the value of the proton
energy, the two components can vary quadratically and cubically  
respectively, while the proton
synchrotron increases only linearly. This means that the relative
contribution of the two photo-processes becomes progressively
more significant.
Furthermore, the contribution of photon-photon cascading
becomes important in redistributing the photon luminosity,
thus complicating the spectral shape. Finally, above
some critical proton luminosity, the system becomes 
supercritical and, as a result,
electrons and radiation rise in an auto-regulatory way 
causing non-linear proton
energy losses (see Fig.~\ref{deltalp} and ~\ref{injectlp}).

This is a very charactersitic property of the hadronic systems,
which can lead it to high efficiencies. 
A detailed study of the system in the
supercritical regime is left for
a forthcoming publication. Preliminary calculations show
that there are at least two loops operating
that can very efficiently extract energy from the protons
and give it to leptons and photons, namely the
proton-synchrotron loop at low proton energies (KM92)
and the automatic photon quenching \citep{stawarzkirk,
petromasti11} at higher ones. The latter loop was
studied analytically by \cite{petromasti12}. Furthermore, it was shown 
that for typical blazar parameters this loop can lead to adoption of
high Doppler factors for the outflow $\delta$; lower values of $\delta$
imply higher proton luminosities leading the system to supercriticality
and to an overproduction of X-rays \citep{PetroMastich2012b}.

The above results generally hold in the case where
the monoenergetic proton distribution is replaced by the
more astrophysically relevant power law (\S 5). In this case,
the maximum energy of proton injection becomes an important
parameter. In addition, results are also sensitive to the 
choice of the proton spectral slope. Interestingly enough,
the calculated neutrino flux is flatter than the producing
proton power law. This might be relevant to
neutrino experiments, like Ice Cube.

In the present paper we have focussed on the case
where the soft photons required as targets  
for photopair and photopion interactions are provided 
from the synchrotron radiation of protons themselves.
While this might not be a realistic case for some applications 
 -- for example, in the case where there is 
a co-accelerated leptonic component, the radiated
synchrotron photons  will be in direct competition
with the  proton
synchrotron ones for the hadronic
interactions -- this approach allowed us to investigate the
time variability in an analogous way to the more familiar
synchrotron-SSC coupling of leptonic models. 
We found (\S 6) that, while the proton synchrotron part always
remains linear to the proton injection, the synchrotron
radiation from photopair, which peaks at higher energies,
 varies quadratically with it.
Thus the hadronic models can reproduce the quadratic 
behaviour of the SSC leptonic models and do so for the
exact same reasons. Perhaps more interesting is that, under certain circumstances,  the 
photopion component that peaks at even higher energies
(see \S 4) shows a cubic behaviour with respect to the
proton injection and therefore to the proton synchrotron
component. This is a unique feature of the hadronic models;
if also considers
the flaring behaviour exhibited when these systems
enter the supercritical regime, 
then it is clear that their temporal behaviour can be quite rich.

In conclusion, we would like to emphasise two key issues
in our approach which, in our opinion, improve the 
concept of hadronic modelling.
(i) The energy conserving scheme that is introduced through
the kinetic equation approach can successfully treat 
the inherent non-linearity of the system. Therefore, once the
initial conditions are chosen, the solution of the system can 
lead either to a subcritical linear solution or to 
a supercritical nonlinear one that, occasionally, might
be oscillatory. There is no way that the
behaviour of the system can be determined a-priori. Therefore, 
time-independent approaches (e.g., adopting a ready
distribution function for the protons) might entirely miss this 
point and lead to erroneous results.
(ii) The careful modelling of photopair and photopion
is crucial.     
We have found that the omission of photopair production can also
lead to erroneous results -- even in cases where the choice of 
initial parameters do not seem to favour it -- when the system settles 
in a state where the internally produced soft photons from photopair 
production are non-negligible. Therefore,
models that do not include it suffer a severe drawback.

Overall, the code presented here can be a useful tool in
examining the one-zone hadronic models of active 
galactic nuclei and gamma ray bursts -- the latter
only in the case where the magnetic field is assumed to be 
low enough not to cause substantial synchrotron losses
to pions and muons before they decay.
It can simultaneously calculate
the photon and the neutrino fluxes from sources with the
only assumption that relativistic protons are injected there.
It takes both photopair and photopion production into account with 
high accuracy, which are very difficult to model.
Finally the energy-conserving scheme of our approach
ensures that we can consider the hadronic models as dynamical
systems and study their temporal behaviour.

\begin{acknowledgements} 
We would like to thank Maria Petropoulou for insightful discussions, and
an anonymous referee whose comments helped improve the manuscript. 
AR acknowledges support by Marie Curie IRG grant 248037 within the FP7 Programme.
\end{acknowledgements}

\bibliographystyle{aa} % style aa.bst
\bibliography{dmpr_20120830} % your references Yourfile.bib
\end{document}